\documentclass[11pt,a4]{article}
\pdfoutput=1
\usepackage{graphicx,amssymb,amsmath,amsfonts}
\usepackage{hyperref}
\usepackage[margin=1.2in]{geometry}
\usepackage[noadjust]{cite}
\usepackage[utf8]{inputenc}

\def\be{\begin{equation}}
\def\ee{\end{equation}}

\newcommand{\alp}{\ensuremath{\alpha^\prime}}

\newcommand{\ssb}{s\bar{s}}
\newcommand{\ccb}{c\bar{c}}
\newcommand{\bbb}{b\bar{b}}
\newcommand{\lclc}{\Lambda_c\overline{\Lambda}_c}

\newcommand{\plm}{\ensuremath{\pm}}

\mathchardef\mhyphenc="2D

\newcommand{\MEV}{\,\mathrm{MeV}\,}
\newcommand{\GEVm}{\,\mathrm{GeV}^{-2}\,}
\newcommand{\GEV}{GeV\(^{-2}\) }

\begin{document}
\begin{titlepage}
\title{\textbf{A tetraquark or not a tetraquark?\\A holography inspired stringy hadron (HISH) perspective}}

\author{\textbf{Jacob Sonnenschein} \\ \href{mailto:cobi@post.tau.ac.il}{cobi@post.tau.ac.il} \and \textbf{Dorin Weissman} \\ \href{mailto:dorinw@mail.tau.ac.il}{dorinw@mail.tau.ac.il}}

\date{\emph{The Raymond and Beverly Sackler School of Physics and Astronomy},\\
	\emph{Tel Aviv University, Ramat Aviv 69978, Israel} \\ \today}
	
\maketitle

\begin{abstract}
We suggest to use  the  state $Y(4630)$, which  decays predominantly to $\Lambda_c\overline\Lambda_c$, as a window to the landscape of tetraquarks. We propose a simple criterion to decide whether a state is a stringy exotic hadron - a tetraquark - or a ``molecule''. If it is the former it should be on a (modified) Regge trajectory. We present the predictions of the mass and width of the higher excited states on the $Y(4630)$ trajectory. We argue that there should exist an analogous $Y_b$ state that decays to $\Lambda_b \overline \Lambda_b$ and describe its trajectory. We conjecture also a similar trajectory for tetraquarks containing strange quarks, and the modified Regge trajectories can in fact be predicted for any resonances found decaying to a baryon-antibaryon pair. En route to the results regarding tetraquarks, we also make some additional predictions on higher excited charmonium states. We briefly discuss the zoo of exotic stringy hadrons and in particular we sketch all the possibilities of tetraquark states.
\end{abstract}
\end{titlepage}

\tableofcontents

\flushbottom

\section {Introduction}
As early as the early days of the quark model and the ``old duality", the question of whether nature exhibits exotic hadronic states on top of the mesons, baryons, and glueballs was raised \cite{Rosner:1968si}, and it has remained unresolved in fact until these days. There has been indirect evidence for the existence of tetraquarks, for instance, it was pointed early on that duality requires the existence of tetraquarks as intermediate states in baryon-antibaryon scattering \cite{Rosner:1968si,Rosner:1974sn}. However, there is no consensual direct observation of a tetraquark state in the hadronic spectrum. In recent years various experimental observations reignited this subject (for reviews see for instance \cite{Krokovny,Esposito:2014rxa}). In particular a remarkable proliferation of exotic charmonium-like resonances has been discovered, some of which, like the charged charmonium-like \(Z_c\) resonances, cannot be interpreted as mesons. Up to date there is no theoretical consensus regarding all the newly discovered states and several conflicting descriptions have been proposed \cite{Esposito:2014rxa}.  

In this note we consider exotic hadrons in the context of the holography inspired stringy hadron (HISH) model \cite{Sonnenschein:2016pim}. Previously we have addressed by using this model the spectra of mesons \cite{Sonnenschein:2014jwa}, baryons \cite{Sonnenschein:2014bia}, and glueballs \cite{Sonnenschein:2015zaa}. We found that a significant part of the known hadrons fall into modified Regge trajectories (MRT) associated with the model. The present work deals with the possibility to have exotic hadronic resonances in this model. We sketch possible arrangements constructed from the basic building blocks, which are open strings, massive particles (quarks) at the endpoints of the open strings, baryonic vertices, diquarks attached to baryonic vertices, and closed strings. We discuss first configurations without quarks where all the strings end on baryonic vertices (BVs) and anti-BVs. These includes the cube, the planar polygon of any even order, tiling the whole plane with hexagons, and more. We also describe setups that include quarks and diquarks in addition to the vertices.

A special class of HISH exotics is that of a baryonic vertex with a diquark attached to it connected by a string to a antibaryonic vertex  with an antidiquark.
This forms what we refer to as a \emph{stringy tetraquark}. For five quark flavors, there is a total of 225 possible tetraquarks, which can be grouped into 15 symmetric, 110 semisymmetric, and 100 asymmetric tetraquarks. 

Classically the stringy tetraquark is described by a string that may rotate or not, and with massive endpoints. This is the same as the string that is associated with a meson or a baryon. The rotating classical strings with  massive endpoints have states which are members of modified Regge trajectories with the same slope as those of the mesons and baryons. Quantum mechanically there are differences between the original Regge trajectories when there are no massive endpoints and the modified trajectories when the strings end on massive particles \cite{Sonnenschein:2016pim}. We will argue that in addition there is a difference between the quantum trajectory of a meson or a baryon and that of a tetraquark, following the same reasoning as for the difference between mesons and baryons. In the later case there is a difference between the two intercepts which is in charge of the difference between the masses of a baryon and a meson with the same orbital angular momentum and endpoint masses. We will suggest a scenario  for which  it is plausible that the difference between the intercept of a tetraquark and a meson will be twice the difference between a baryon and meson. 
  
The most natural decay mode of the tetraquark configuration, at least for high spin excited states, where the description of the tetraquark as a diquark and antidiquark should hold, is similar to that of meson or a baryon. It involves the breaking of the string into two strings, one connecting the BV to a quark and the other an antiquark to an anti-BV (see figure \ref{holdecay}). This means that the natural decay mode of the stringy tetraquark, provided it is sufficiently massive, is to a baryon-antibaryon pair, as a BV and an anti-BV are already parts of the tetraquark. The computation of the decay width of such a process is in fact very similar to that of a meson decaying into two mesons \cite{Peeters:2005fq}.

As was mentioned above an important question about the peculiar states that have been observed recently is whether they are genuine exotic hadrons or ``molecules'' \cite{Karliner:2015ina}, i.e. bound states of non-exotic hadrons. We propose a simple criterion to help decide whether a state is a genuine stringy exotic hadron or a molecule. If it is the former it should be on a (modified) Regge trajectory while for the latter there is no apparent reason to have such a trajectory. In QCD terms, this is the difference between the color interaction of the diquark and antidiquark (equivalent to an antiquark and a quark) and the mechanism, probably pion exchange \cite{Karliner:2015ina}, holding a bound state of hadrons together. It should be emphasized however that the stringy description is expected to be valid for excited states, when long strings are formed between the diquark and antidiquark forming the tetraquark. It is not obvious that the lower tetraquark states (the ones that we are more likely to see in experiment) should be on trajectories, but experience has shown us that this holds for both mesons and baryons, even those made up of heavy quarks \cite{Sonnenschein:2014jwa,Sonnenschein:2014bia}. The notion that the tetraquarks have Regge behavior is not new, dating back to the 1970s \cite{Rossi:1977cy,Montanet:1980te}, but now, with the wealth of new data, it is time to reexamine it. 

Out of the zoo of special states we suggest to use the state $Y(4630)$, which  decays predominantly to $\lclc$, as a window to the landscape of tetraquarks. The baryon-antibaryon decay signals it out as a state that can match our description of a stringy tetraquark. We propose to use our  simple criterion to decide whether this state is a genuine stringy exotic hadron or a molecule. We present the predictions of the mass and width of the higher excited states of the $Y(4630)$ trajectory. We compute the trajectories associated with  $M^2$ and the angular momentum  $J$, as well as the trajectories associated with $M^2$ and the excitation level $n$. In this way we also make some predictions of yet undiscovered higher excited charmonium states.

We argue that there should exist an analogous $Y_b$ state in the bottomonium sector that decays to $\Lambda_b\overline{\Lambda}_b$, which should also belong to a trajectory. Whereas such a state can follow also from a non-stringy picture,  the existence of a (modified) trajectory would be clearly an indication of a stringy nature.  We also propose a similar trajectory for tetraquarks containing \(\ssb\) and we would like to encourage an experimental search for such tetraquarks decaying to $\Lambda\overline\Lambda$. A priori such objects made out of light quarks are also not excluded, and in fact, any resonance which may be found to decay to any pair of baryon and antibaryon - such as \(\Lambda_c\overline\Sigma_c\), \(\Lambda_c\overline\Lambda_b\), \(\Lambda_b\overline\Lambda\), or many other options - could be a tetraquark with an appropriate trajectory following from its stringy structure.

The paper is organized as follows. In section \ref{sec:HISH} we briefly review the basic concepts and features of the HISH model. In particular we explain the two main ingredients of the model, the string endpoint mass and the baryonic vertex. We also describe the basic string configurations that constitute the mesons, baryons, and glueballs. In section \ref{sec:exoticHISH} we discuss possible exotic stringy configurations. We mention several classes of them, for instance configurations with no quarks or diquarks, formed of only baryonic and antibaryonic vertices. We also specify the classes of exotic tetraquark configurations. In section \ref{sec:tetra} we describe the symmetric, semisymmetric and asymmetric tetraquark configurations. We then describe in section \ref{sec:tetra_traj} the modified Regge trajectories of tetraquarks. In particular we discuss the differences between the intercept of mesons and baryons and we conjecture about the intercept of the exotic tetraquarks. Section \ref{sec:decays} is devoted to the decay processes of the HISH tetra quarks. We write down the width of the decay of a tetraquark which has in fact the same form as the width of the decay of a meson and a baryon.

Readers interested in the phenomenological aspects of our work and predictions of new exotic states will find them in section \ref{sec:pheno}, where we make the comparison between the HISH tetraquarks and hadron phenomenology. We start with the tetraquark in the charmonium sector. We show that all of the known charmonium states - i.e. the non-exotic \(\ccb\) mesons - fit nicely on Regge trajectories, and extract the parameters of the Regge slope and endpoint mass of the \(c\) quark. We review the $Y(4630)$ resonance which is our main candidate to be a tetraquark. We describe the various options to understand this state and we write down its modified Regge trajectory. We then describe the possibility that there are such tetraquarks also in the bottomonium sector and in the light quark sector, and offer some tentative predictions for these states. In section \ref{sec:summary} we summarize and mention several open questions. 

\section{A brief review of  the HISH model} \label{sec:HISH}
The holographic duality is an equivalence between certain bulk string theories and boundary field theories. The original duality was between the ${\cal N}=4$ SYM theory and string theory in $AdS_5\times S^5$. Obviously the  ${\cal N}=4$ theory is not the right framework to describe hadrons that resemble those found in nature. Instead we need a stringy dual of a four dimensional gauge dynamical system which is  non-supersymmetric, non-conformal, and confining. The  two main requirements on the desired string background is that it admits confining Wilson lines, and that it is dual to a boundary that  includes a matter sector, which is invariant under a chiral flavor symmetry that is spontaneously broken. There are by now  several ways to get a string background which is dual to a confining boundary field theory. For a review paper and a list of relevant references see for instance   \cite{Aharony:2002up}.

Practically most of the applications of holography of both conformal and non-conformal  backgrounds are based on relating bulk fields (not strings) and operators on the dual  boundary field theory. This is based on the usual limit of $\alp\rightarrow 0$  with which we go, for instance, from a closed string theory to a theory of gravity.

However, to describe realistic hadrons we need strings rather than bulk fields since in nature the string tension, which is inversely proportional to $\alp$, is not very large. In gauge dynamical terms the IR region is characterized by $\lambda= g^2 N_c$ of order one and not a very large one.

It is well known that in holography there is a wide sector of hadronic physical observables  which cannot be faithfully described by bulk fields but rather require dual stringy phenomena. This is the case for a Wilson, 't Hooft, or Polyakov line.

In the  holography inspired stringy hadron (HISH) model  we argue that in fact also  the  spectra, decays and other properties of hadrons - mesons, baryons, glueballs and exotics - should be recast only by holographic stringy configurations and not by fields. The major  argument against describing the hadron spectra in terms of fluctuations of fields, like bulk fields or modes on probe flavor branes, is that they generically do not admit Regge trajectories in the spectra, namely, the (almost) linear relation between $M^2$ and the angular momentum $J$. Moreover, for top-down models with the assignment of mesons as fluctuations of flavor branes  one can get mesons only with 
 $J=0$ or $J=1$. Higher $J$ will have to be related to strings, but then there is a big gap of order $\lambda$ (or some fractional power of $\lambda$ depending on the model) between the low and  high $J$ mesons. Similarly  the attempts to get the observed  linearity between $M^2$     and the excitation number $n$ are problematic whereas for strings it is an obvious property.

The construction of the HISH model is based on the following steps. (i) Analyzing string configurations in holographic string models that correspond to hadrons, (ii) devising a  transition from the holographic regime of large $N_c$ and large $\lambda$ to the real world that bypasses taking the limits of $\frac{1}{N_c}$ and $\frac{1}{\lambda}$ expansions, (iii)  proposing a model of stringy hadrons in flat four dimensions that is inspired by the corresponding holographic strings, (iv) confronting the  outcome of the models with the experimental data (as was done in \cite{Sonnenschein:2014jwa,Sonnenschein:2014bia,Sonnenschein:2015zaa}).

Confining holographic models are characterized  by a ``wall" that truncates in one way or another the range of the holographic radial direction. A common feature to all the holographic stringy hadrons is that there is a segment of the string that stretches along a constant radial coordinate in the vicinity of the ``wall", as in figure \ref{tetrahol}. For a stringy glueball it is the whole folded closed string that rotates there and for an open string it is part of the string, the horizontal segment, that connects with vertical segments either to the boundary for a Wilson line or to flavor branes for a meson or baryon. The fact that the classical solutions of the flatly  rotating strings reside at fixed radial direction is a main rationale behind the map between rotating strings in curved spacetime and rotating strings in flat spacetime. A key ingredient of the map is the ``string endpoint mass'', $m_{sep}$, that provides in the flat spacetime description the dual of the vertical string segments. It is important to note that this mass parameter is neither the QCD mass parameter nor that of the constituent quark mass, and that the $m_{sep}$ parameter is not an exact map of a vertical segment. The endpoint mass is obtained as an approximation that is more accurate the longer the horizontal string, when the vertical segments are most pronounced (for shorter strings the configuration will resemble a more curved ``U'' shape).

The stringy picture of mesons has been thoroughly investigated in the past (see \cite{Collins:book} and references therein). It turns out that also in recent years there were several attempts to describe hadrons in terms of strings. Papers on the subject that have certain overlap with our approach are \cite{Baker:2002km,Schreiber:2004ie,Hellerman:2013kba,Zahn:2013yma}. Another approach to the stringy nature of QCD is the approach of low-energy effective theory on long strings. This  approach is similar to  the approach presented in this paper. A recent review of the subject can be found in \cite{Aharony:2013ipa}.

The HISH model describes hadrons in terms of the following  basic ingredients.
\begin{itemize}
\item
Open strings  which are characterized by a tension $T$ (or a slope $\alp=\frac{1}{2\pi T}$). The open string generically has a given energy (mass) and angular momentum associated with its  rotation. The latter gets contribution from the classical configuration and in addition there is also a quantum contribution (the intercept $a$).  Any such open string can be in its ground state or in an excited state.
\item
Massive particles - or ``quarks'' - attached to the ends of the open strings which can have four\footnote{There is no a priori reason to impose \(m_u = m_d\) in our model, but, for all our purposes the two masses can be taken to be the same. The difference between them is smaller than the degree to which the masses can be determined from the phenomenology.} different values $m_{u/d}$, $m_s$, $m_c$, $m_b$. The latter are determined by fitting the experimental spectra of hadrons. These particles naturally contribute to the energy and angular momentum of the hadron of which they are part.
\item
``Baryonic vertices'' (BV) which are connected to a net number of three (\(N_c\)) strings. One particularly important configuration would involve a single long string and two very short ones. Since the endpoints of the two short strings are one next to the other we can consider them as forming a diquark.  We emphasize that unlike other models which have quarks and diquarks as elementary particles (such as \cite{Friedmann:2009mx}), in the HISH model the diquark is always attached to a baryonic vertex. A BV can be also connected to a combination of quarks and antibaryonic vertices as will be discussed in the next section.  
\item
  Closed strings which have an effective tension that is twice the tension of the open string, leading to a slope that is halved, $\alp_{closed}= \frac12 \alp_{open}$. They can have non-trivial angular momentum by taking a configuration of a folded rotating string. The excitation number of a closed string is necessarily even, and so is the angular momentum (on the leading Regge trajectory). Their intercept, the quantum contribution to the angular momentum, is also twice that of an open string ($a_{closed}=  2 a_{open}$).
	\end{itemize}
	
Hadrons, namely mesons, baryons and glueballs are being constructed in the simplest manner from the HISH building blocks.
 \begin{itemize}
\item
A single open string attached to two massive endpoints corresponds to a meson.
\item
A single string that  connects on one hand to a quark and on the other hand a baryonic vertex with a diquark attached to it is the HISH description of a baryon.
\item
A single closed string is a glueball.
\end{itemize}
\section{Exotic hadrons in the HISH model} \label{sec:exoticHISH}
In addition to the basic hadron configurations with which we ended the previous section, there is a very wide range of other options to combine the elementary building blocks of the HISH model.

To decide which of them are stable,  one has to show that they are solutions of the quantum equations of motion. One can check the equations of the HISH model \cite{Sonnenschein:2016pim} or better, check the underlying holographic string equations. The physical properties that enter the former equations are the string tension, the endpoint masses, and the angular momentum (or the corresponding centrifugal force). There are also electromagnetic forces between the endpoint charges, although the impact of those is typically much smaller. These are features of the classical configurations. Quantum mechanically there is a Casimir force effect which is related to the intercept.  For instance the $\eta$ meson has $J=L=S=0$ and it is the negative intercept $(a \sim -0.23)$ that corresponds to a repulsive Casimir force that balances the tension. In addition as will be discussed in section \ref{sec:decays} the main quantum instability associated with the breaking mechanism  of the string. We will not perform the stability analysis for all  the possible HISH states (but just refer to the decays  of  tetraquarks in section \ref{sec:decays}) . Moreover, we do not intend to survey all the  possible  stringy configurations since we will be interested in this work only in those that correspond to tetraquarks. Nevertheless, we sketch here several classes of possible configurations.

Configurations that involve open strings are characterized by $n_B$, $n_{\bar B}$, $n_q$, and $n_{\bar q}$, which are the numbers of baryonic vertices, antibaryonic vertices, quarks, and antiquarks respectively. They must obey the relation
\be \frac{1}{3}(n_B- n_{\bar B}) - (n_q- n_{\bar q}) = 0 \ee

\begin{figure*} \centering
	\includegraphics[width=0.75\textwidth]{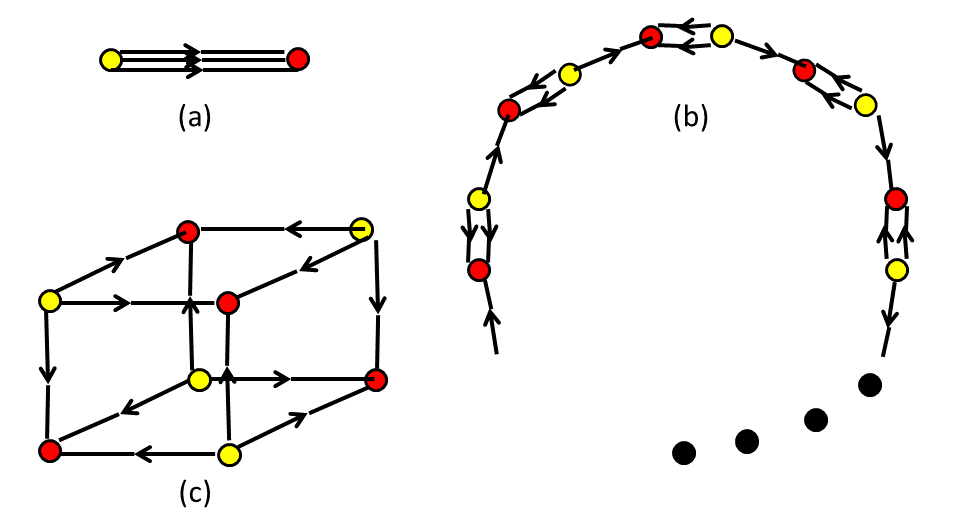}
				\caption{\label{bvsantibvs} Configurations of BVs and anti-BVs. (a) Three strings connecting one BV and one anti-BV. (b) A planar polygon.  (c) A cube.}
			\end{figure*}

\paragraph{Exotic configurations without quarks:} For the case of no quarks or antiquarks, namely with only baryonic vertices and antibaryonic vertices, we necessarily have $n_B= n_{\bar B}$. In this case we can have
a configuration similar to a closed string with three strings connecting the BV and the anti-BV. See (a) in figure (\ref{bvsantibvs}).

\begin{figure} \centering
	\includegraphics[width=0.48\textwidth]{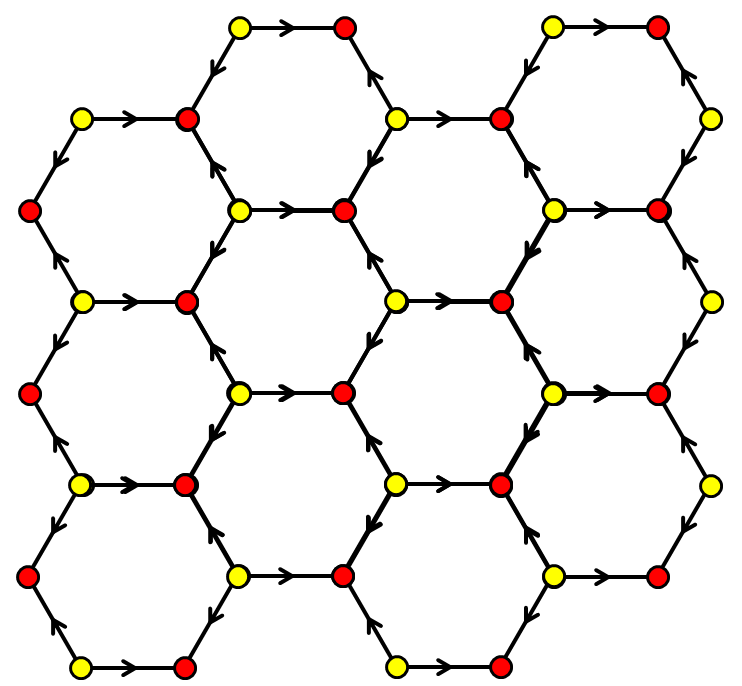}
				\caption{\label{hexagon} Tiling the whole plane with hexagons of BVs and anti-BVs. The pattern can be repeated to infinity.}
			\end{figure}

			Another class of possible configurations consists of any planar polygon with an even number $n_B + n_{\bar B}$ vertices such that from each baryonic vertex there is one line going to a neighboring antibaryonic vertex, and two lines going to the other anti-BV on the other side, as is depicted in figure \ref{bvsantibvs} (b). We can have also have a three dimensional configuration such as the cube, where each vertex has three strings connected to it. This is (c) in figure \ref{bvsantibvs}. In addition we can tile the whole plane with hexagons, as seen in figure \ref{hexagon}.

Let us  emphasize again that we do not aim here at finding all possible configurations and not to check their stability. We merely demonstrate several potential possibilities, and we do this only for the case of \(N_c = 3\). The number of strings can be further generalized to give even more possibilities.

\paragraph{Exotic hadrons with quarks:} Next we can discuss the case of quarks and antiquarks on top of the BVs and anti-BVs.

The basic configuration of a baryonic vertex is that there are three strings attached to it. As was mentioned above, baryons are in fact characterized by a string tension, or a slope which is the same as for mesons, and hence one concludes that there is only one long string attached to the BV, with two short strings connected it as a diquark. 

\begin{figure*} \centering
	\includegraphics[width=0.48\textwidth]{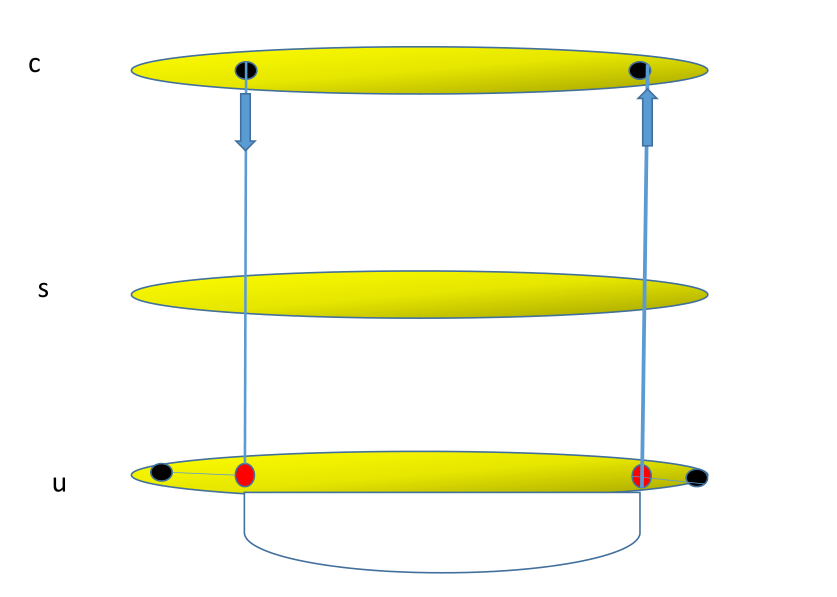}
				\caption{\label{tetrahol} An example of a  holographic tetraquark based on a diquark of $c$ and $u$ and an antidiquark of $\bar c$ and $\bar u$}
			\end{figure*}

Generically, the single string  can be connected either to another quark or to an antibaryonic vertex to which an antidiquark is connected. The first possibility is the baryon, while and the second one is what we will refer to as the tetraquark. In figure \ref{tetrahol} we draw for example a tetraquark including $c$ and $\bar c$ quarks. We distinguish between a baryonic vertex and an antibaryonic vertex by the orientation of the strings attached to it. We take that strings are coming into the BV and are coming out of the anti-BV.

\begin{figure*} \centering
	\includegraphics[width=0.75\textwidth]{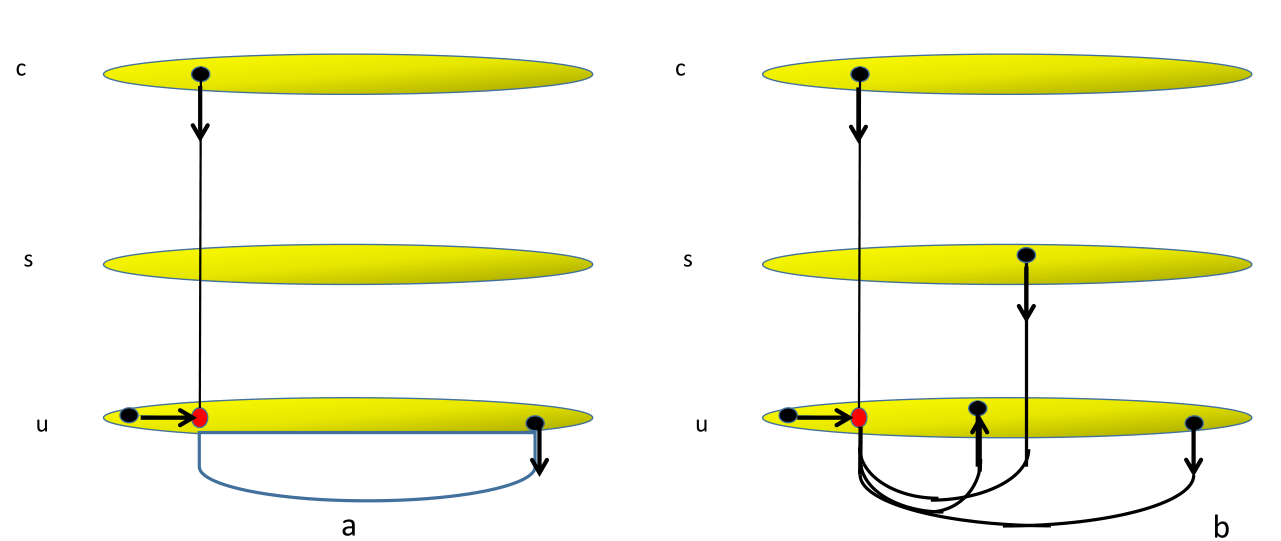}
				\caption{\label{molecule} The holographic picture of $\Lambda_c$ baryon (a) and a pentaquark (b) with an added quark and an antiquark.}
			\end{figure*}

In fact the condition for the neutrality of Abelian gauge field that resides on the wrapped $D$-brane that constitutes the baryonic vertex \cite{Witten:1998zw}, can be obeyed also if there are $N_c+ k$ strings and $k$ anti-strings for the general case of $N_c$ \cite{Brandhuber:1998xy}. In particular we can have a BV with an attached diquark and single string whose end is a quark and, in addition, $k$ strings that end on a quark and $k$ strings with the opposite orientation that end on antiquarks. As such we get a configuration which can be considered as pentaquarks (for \(k = 1\)) or ``heptaquarks'' etc. for higher \(k\). In figure \ref{molecule} we draw the holographic picture of a pentaquark that has the same quark content of a $\Lambda_c$ and a $K$ meson. It is easy to notice that these exotic hadrons may have electric charges that cannot be carried by an ordinary hadron, in particular there may be a pentaquark state with electric charge $3$.

Based on the experience with ordinary baryons that are based on a single string it is probable that the additional pairs of string-antistring will connect the BV to the nearby flavor branes with very short strings like for the diquarks. One can similarly attach quark-antiquark pairs to the tetraquark configuration of above thus yielding hexaquarks, octoquarks, and so forth.

\begin{figure} \centering
	\includegraphics[width=0.36\textwidth]{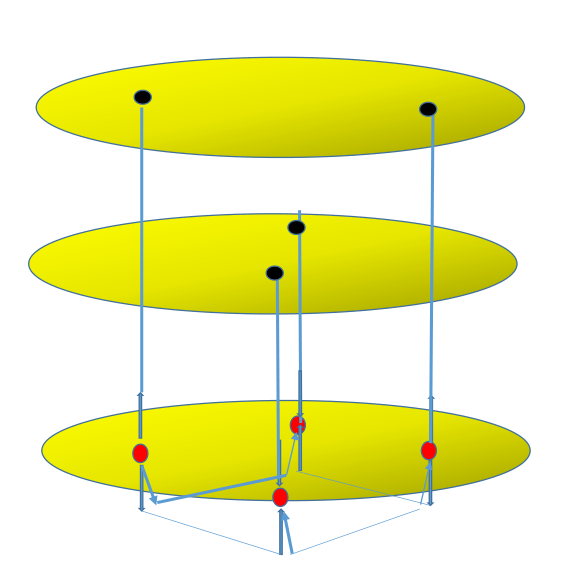}
				\caption{\label{multiquark} An example of a holographic tetraquark built with two baryonic and two antibaryonic vertices.}
			\end{figure}

Instead of a single BV or a pair of one BV and one anti-BV as was just described one can also have configurations where $n_B$ baryonic vertices are connected to $n_{\bar B} = n_B$ antibaryonic vertices. See for instance a configuration with $n_B=n_{\bar B}=2$ drawn in figure \ref{multiquark}. In this case the exotic configuration can carry a charge of  $-2$, $-1$, $0$, $1$ or $2$.

			In these configurations each BV is connected to two anti-BVs and to a quark, and the anti-BVs similarly connect each to two BVs and an antiquarks.
 
One can also imagine the configurations described above where the pairs of strings with opposite orientations, are not connected to flavor branes in the holographic picture or to a quark and antiquark in the HISH picture, but rather merge one into the other thus forming closed strings.

This is just a partial list of possible hypothetical hadronic configurations. In a future publication we will discuss these exotics in more detail.

\section{Tetraquark configurations} \label{sec:tetra}
From the zoo of potential exotic multiquark states discussed in the previous section we now want to focus on the tetraquarks. The configuration of a BV attached to a diquark connected by a string to an anti-BV attached to an antidiquark, which is referred to as tetraquark, can be classified by distinguishing between the following possibilities.
\begin{itemize}
\item Symmetric tetraquarks where the antidiquark is built from the antiquarks associated with those found in the diquark. For instance, in figure (\ref{tetrahol}) the diquark $cu$, and the antidiquark $\overline {cu}$.
\item Semisymmetric tetraquarks in which there is one pair of quark and antiquark of the same flavor and one pair which include a quark and an antiquark of different flavors, for instance \((cu)(\overline{cs})\).
\item Asymmetric tetraquarks, where both pairs are of different flavor.
\end{itemize}
\subsection{Symmetric tetraquarks}
\begin{figure*} \centering
	\includegraphics[width=0.95\textwidth]{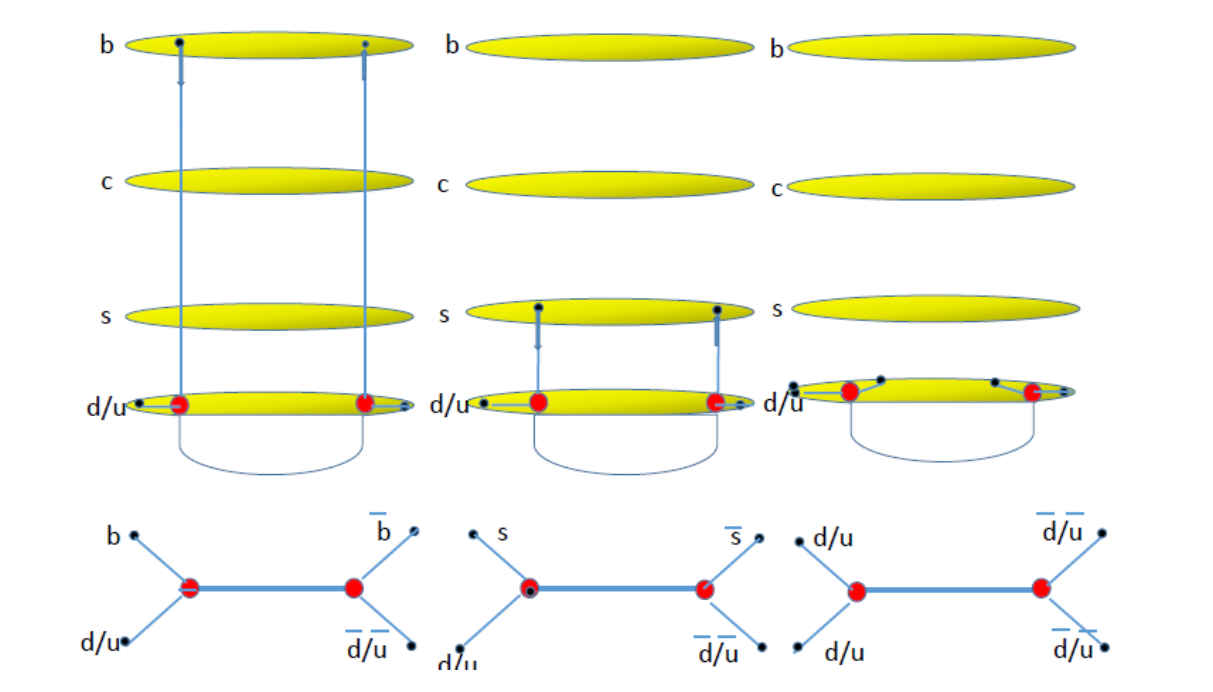}
			\caption{\label{symtetra} Symmetric holographic tetraquarks and their corresponding HISH pictures. The diquark  include a $b$, $s$ and $u/d$ quarks for the left, middle and right templates correspondingly.}
			\end{figure*}

Let us start with a description the configurations described above in holography. Similar to figure \ref{tetrahol} where the diquark includes a $c$ quark, one can have  also  configurations that include  instead  a diquark with any other flavor quarks, as shown in figure \ref{symtetra}.

Notice that these configurations are symmetric in the sense that for every quark connected to the baryonic vertex there is an antiquark of the same flavor connected to the antibaryonic vertex. It thus obvious that these symmetric tetraquark configurations are flavorless and carry zero electric charge. In the lower part of figure \ref{symtetra} we  draw the corresponding pictures of each of the tetraquarks in the HISH setup. In fact, we drew only the cases where one of quarks that build the diquarks is a light $u$ or $d$ quarks but in fact we can also have other combinations like those with a $c$ and $s$ diquark. Altogether there are 15 different symmetric tetraquarks for 15 unique types of diquarks composed of quarks of five flavors (as the diquark \((ij)\) is identical to the diquark \((ji)\)).

\subsection{Semisymmetric tetraquarks} 
Another class of tetraquark is the semisymmetric one. In this class of tetraquarks only one pair in matched and the other is not. Thus the flavor content of these exotic hadrons is the same as of mesons that carry non-trivial flavor. We have 5 possibilities for a matched pair and 20 for the unmatched pair, so altogether there are 100 possible semisymmetric tetraquarks. For example in  figure \ref{asymtetra} the left figure may describe a semisymmetric tetraquark with a matched pair of \(u\) and \(\overline u\), with a \(b\) and a \(\overline c\) as the unmatched pair.

Unlike the symmetric tetraquarks which are always chargeless, semisymmetric tetraquarks can carry a charge of $+1$, 0, or $+1$.

\subsection{Asymmetric tetraquarks}
 We could have any pair of quark forming the diquark and any two antiquarks forming the antidiquarks thus altogether there are a priori $15\times 15=225$ possibilities of tetraquarks. Out of the 225 tetraquarks, we have 15 that are symmetric, 100 semisymmetric, and thus 110 are asymmetric ones. The asymmetric tetraquarks can carry a charge of $-2$, $-1$ , 0, $+1$, or $+2$, with a charge of \(\pm2\) being an obviously exotic feature (for hadrons with baryon number zero).

\begin{figure*} \centering
	\includegraphics[width=0.95\textwidth]{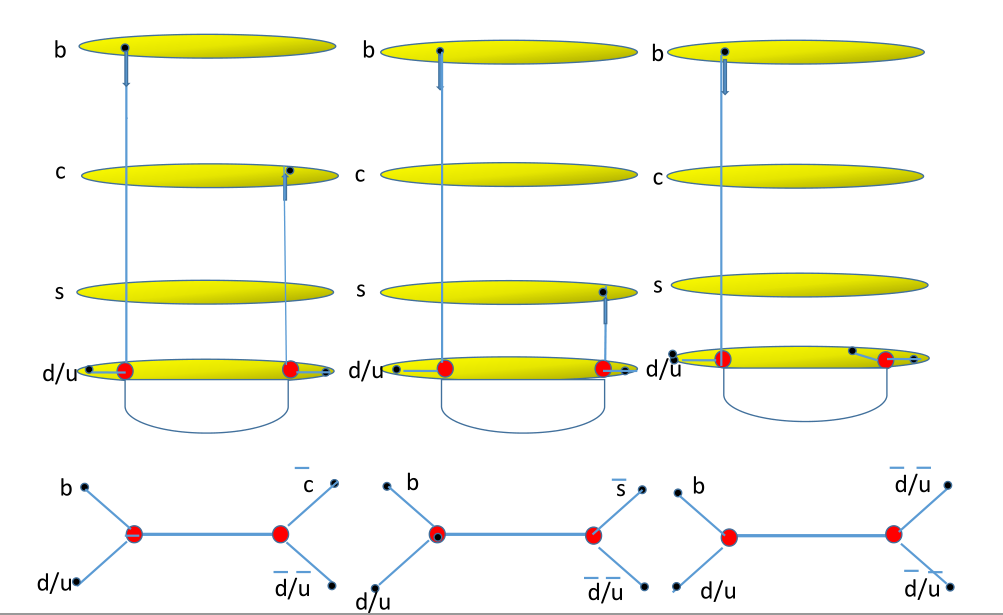}
				\caption{\label{asymtetra} Semisymmetric and asymmetric holographic tetraquarks built from a diquark that include a $b$ quark  and their corresponding HISH pictures.}
			\end{figure*}

In figure \ref{asymtetra}, the left figure is taken as $(bu)(\overline{cd})$, an example of an asymmetric tetraquark.

\section{The modified Regge trajectories of tetraquarks} \label{sec:tetra_traj}
Following the holographic description, the four dimensional flat spacetime HISH description of a tetraquark takes the form of a single string with massive endpoints, which in addition to mass carry certain flavor and charge. It is thus clear that similarly to the HISH configurations that describe mesons and baryons, the tetraquarks also admit modified Regge trajectories (MRT). As was analyzed in \cite{Sonnenschein:2014jwa} the modification of  the linear Regge trajectories is due to the massive particles at the endpoints of the string.\footnote{Additional modification is caused by the electric charges of the endpoints. This modification is less significant than that due to the masses so we will ignore it in what follows.}

The endpoint masses are given, in the mapping from holographic to flat spacetime, by the length of the vertical segments of the string. For a diquark, there is a priori a contribution to the endpoint mass from the baryonic vertex. However, in \cite{Sonnenschein:2014bia} it was found that the BV and the short strings completing the diquark do not make a significant contribution to the endpoint mass, at least not in the case where one or both of the quarks in the diquark is light, i.e. a \(d\) or \(u\) quark. The mass of the BV also depends on its height in the radial coordinate, in a way that will vary between different models, and in the case of a diquark composed of a heavy quark and a light one, it is energetically favorable for the BV to remain on the lower flavor brane, where it is less massive. The details will be model dependent. With these two points in mind, but in particularly the phenomenological result of \cite{Sonnenschein:2014bia}, we can take the endpoint mass of a \(cu\) diquark, for example, to be simply the mass of the \(c\) quark.

The existence of the baryonic vertex in the holographic picture may affect the intercept of the HISH in the quantum case, as will be discussed in section \ref{sec:quantum_traj}.

\subsection{Classical trajectories}
The trajectories are given in terms of the energy and the angular momentum of the string with massive endpoints. In the holographic picture the masses of the endpoints come from the vertical segments of the string and the BV in the case of a diquark, but from a flat spacetime perspective we do not see the details, only the masses themselves. Classically the trajectories are given by 
\be E = \sum_{i=1,2}m_i\left(\frac{\beta_i\arcsin(\beta_i)+\sqrt{1-\beta_i^2}}{1-\beta_i^2}\right) \label{eq:massGenE} \,,\ee
		\be J = \sum_{i=1,2}\pi\alp m_i^2\frac{\beta_i^2}{(1-\beta_i^2)^2}\left(\arcsin(\beta_i)+\beta_i\sqrt{1-\beta_i^2}\right) \label{eq:massGenJ} \,,\ee
where $\beta_i= \omega l_i$ is the velocity of an endpoint, and $l_i$ is the length of each of the  arms of the string, namely the segments from the center of mass, around which the string rotates, to the endpoints. The velocities are related by the boundary condition
 \be \frac{T}{\omega} = m_1\frac{\beta_1}{1-\beta_1^2} = m_2\frac{\beta_2}{1-\beta_2^2} \label{eq:boundaryGen}\,.\ee

	In the low mass limit where the endpoints move at a speed close to the speed of light, \(\beta_i \rightarrow 1\), we have an expansion in \((m/E)\):
		\be \begin{split} & J = \alp E^2\times \\ &\left(1-\sum_{i=1}^2\left(\frac{4\sqrt{\pi}}{3}\left(\frac{m_i}{E}\right)^{3/2} 
		+ \frac{2\sqrt{\pi^3}}{10\sqrt{2}}\left(\frac{m_i}{E}\right)^{5/2} + \cdots\right)\right)\,, \label{eq:lowMass} \end{split}\ee
		from which we can easily see that the linear Regge behavior is restored in the limit \(m_i\rightarrow 0\), and that the first correction is proportional to \(\sqrt{E}\). 

		The high mass limit holds when
		\be (E-m_1-m_2)/(m_1+m_2) \ll 1\,.\ee
		We take the limit \(\beta_i \rightarrow 0\), and the expansion is
		\be \begin{split} J & = \frac{4\pi}{3\sqrt{3}}\alp \sqrt{\frac{m_1m_2}{m1+m_2}} (E-m_1-m_2)^{3/2} \\
		& + \frac{7\sqrt{2}\pi}{27\sqrt{3}} \alp  \frac{m_1^2-m_1m_2+m_2^2}{m_1m_2\sqrt{(m_1+m_2)^3}} (E-m_1-m_2)^{5/2}
			+ \cdots \label{eq:highMass}\,. \end{split}\ee
\subsection{Quantum trajectories} \label{sec:quantum_traj}
The classical trajectories are uplifted to the quantum trajectories when quantum fluctuations are included. For a linear Regge trajectory of a string with no massive endpoints the passage from the classical to the quantum picture is simply via the replacement
\be
J_{cl}=\alp E_{cl}^2\qquad\rightarrow\qquad  J+ n = \alp E^2 + a
\ee
where $n$ are the eigenvalues of the world sheet Hamiltonian, $w_n=n$, of the $n$-th excitation level, and $a$ is the intercept, $a=\frac{D-2}{2}\sum_n w_n$. In the last expression, $D$ is the number of dimensions in the target spacetime. In fact this expression holds only for the critical dimension and for the case of non-critical dimensions it is corrected by the Liouville term. As was explained in detail in \cite{Sonnenschein:2016pim}, for the MRT the story is much more complicated in various respects.

Firstly, the eigenvalues $w_n$ are affected by the masses at the ends of the string, $w_n = w_n(n, \frac{TL}{m})\neq n$. Moreover \cite{Zahn:2013yma}, the eigenvalues are different for the fluctuations transverse to the plane of rotation and for the one along the direction of the  string (which is not there at all in the massless case). Correspondingly, the intercept $a$ is modified. For the static non-rotating string this was derived in \cite{Lambiase:1995st}. 

Another complication is that since the string associated with the tetraquark resides in four dimensions, i.e. a non-critical dimension, one has to include the Liouville or the Polchinski-Strominger term for consistent quantization \cite{Polchinski:1991ax}. This term diverges for the massless case and has to be regularized and renormalized. This was discussed for the massless case in \cite{Hellerman:2013kba} and was touched upon for the massive case in \cite{Sonnenschein:2016pim}.

The passage from the classical to the quantum trajectory that we have just presented is the one required for the stringy meson. The question is whether the same treatment applies also to the exotic tetraquark configuration, or whether there are special features associated with it. On the one hand it seems that from the point of view of the fluctuations of the string there is no difference between the mesonic, baryonic, and the tetraquark string, since in all these three cases it is a string between massive endpoints. However, that is not the whole story. In fact just by comparing the mesonic and baryonic trajectories we observe, by comparison to the experimental data, that although the tension and the endpoint masses are very similar, there are several differences.

Obviously the angular momenta along the baryon trajectories are half integers whereas for the mesons they are integers. This follows of course from the contribution of the spin degrees of freedom to the total angular momentum. The role of the spin in the HISH has not yet been worked out. In the holographic picture the quark is in fact the vertical (along the holographic radial direction) segment of the string and its endpoint is on the flavor brane. Correspondingly the spin associated with each quark can be related to the string along the radial direction or to its endpoint. The former is the case if the vertical segment is a fermionic string. In this case one needs to explain how to connect the horizontal bosonic string to the vertical fermionic spin-carrying one. The latter option is that the connection point on the flavor brane is a fermion.  In principle one can attribute a spin $S=\frac12$ to the baryonic vertex, however that will not explain the spin values of the various mesons. So we will assume here that there is a spin $S=\frac12$ to each connection of a string and a flavor brane.

To compare the intercept of hadrons with different spin, like baryon and a meson, it is useful to define a different intercept as follows. Let us start with the linear case, which we can write as
\be
J= L\pm S =\alp E^2 + a \qquad \rightarrow \qquad L =  \alp E^2 + \tilde a \ee
by defining
\be \tilde a \equiv a \mp S \ee
For the MRT of strings with massive endpoints  we still define $\tilde a$ in the same way and relate it to the trajectory of $L$, the orbital angular momentum.

When comparing the intercepts of mesons and baryons we find that they are different even when they are composed of the same type of quarks. In particular we find that there is a difference between $\tilde a$ of the lightest baryon composed of $u$ and $d$ quarks and the lightest meson (which is not a Goldstone boson).

As was mentioned before in both cases the intercept is a measure of quantum fluctuations of a straight string between massive particles on its ends. We must conclude that the fluctuations of a system that includes a string and an  endpoint built from a BV and a diquark is different from that with  an ordinary quark as an endpoint. There are two possible reasons for this difference: (i) the quantum fluctuations of the BV, or (ii) a change of the boundary condition for the string  by the BV which changes the eigenvalues its fluctuations.

If we denote by $\tilde a_m$ and $\tilde a_b$ the intercepts of a meson and a baryon of the same flavor structure,  the   the difference of the intercepts is $
\Delta\tilde a =\tilde a_b-  \tilde a_m $. If the cause of the difference is the quantum fluctuations of the BV then we anticipate that 
\be \tilde a_t\equiv \tilde a_m + 2 \Delta \tilde a = \tilde a_b + \Delta \tilde a\,. \label{eq:intercept_tetra}\ee
But the intercept of the tetraquark will not be of this form  if   it is due to the change of the  boundary conditions of the string. In any case, since we do not know how to determine this difference from string theory (recall that the BV is a wrapped $D_p$ brane over a $c_p$ cycle), we will use the experimental data to determine $\tilde a_b$,  $\tilde a_m$,  and $\tilde a_t$.
   
Another difference between the meson and baryon trajectories is the fact that for baryons there is a different intercept for even and odd orbital angular momentum states. This happens for baryons composed of light quarks, namely the \(N\) and \(\Delta\) baryons, and not for those built from heavy quarks. Where this effect occurs we see two parallel trajectories, with the trajectory of the odd states above that of the even states,
\be L = \begin{cases}
	\alpha M^2 + a_{e}\,, & L \text{ even} \\
	\alpha M^2 + a_{o}\,, & L \text{ odd}\,. \\ \end{cases}
\ee
For the \(N\) trajectory \(a_e - a_o \approx 0.4\), while for the \(\Delta\), \(a_e - a_o \approx 1\). The source of this difference, and its magnitude, must depend on the spins involved. In the baryon we have an spin 1/2 quark, and a diquark which can carry spin 0 or 1.

In the tetraquark, we have a diquark and antidiquark, where each may carry spin. If the source of the difference between the even and odd intercepts is the BV and the diquark, we should expect a similar split also for the tetraquark trajectories built from light quarks. If, on the other hand, it occurs only when the total angular momentum is a half-integer, and depends on an interaction between the half-integer spin of the quark and the integer spin of the diquark, then for the tetraquarks that have integer angular momentum we should not have this phenomenon. The details of this phenomenon in the baryon spectrum have not been worked out yet, and a fuller understanding is needed before speculating further on its existence in the case of the tetraquark.

\section{Decays of the tetraquarks} \label{sec:decays}
In a similar manner to the decay of mesons, baryons, and glueballs, the exotic hadrons can also decay generically to a combination of exotic and ordinary hadrons. 

For low mass tetraquarks, the constituent quarks and antiquarks may be close enough to each other to annihilate and thus decay, but for excited tetraquarks that are described by a diquark and antidiquark joined by a (long) string the dominant mechanism of decay is a split of a string and a generation of a quark-antiquark pair at the two ends of the torn apart string. In this way we get a BV connected to a quark and a diquark on one side of the split string - which gives a baryon, while on the other side we get an antibaryon.

\begin{figure} \centering
	\includegraphics[width=0.48\textwidth]{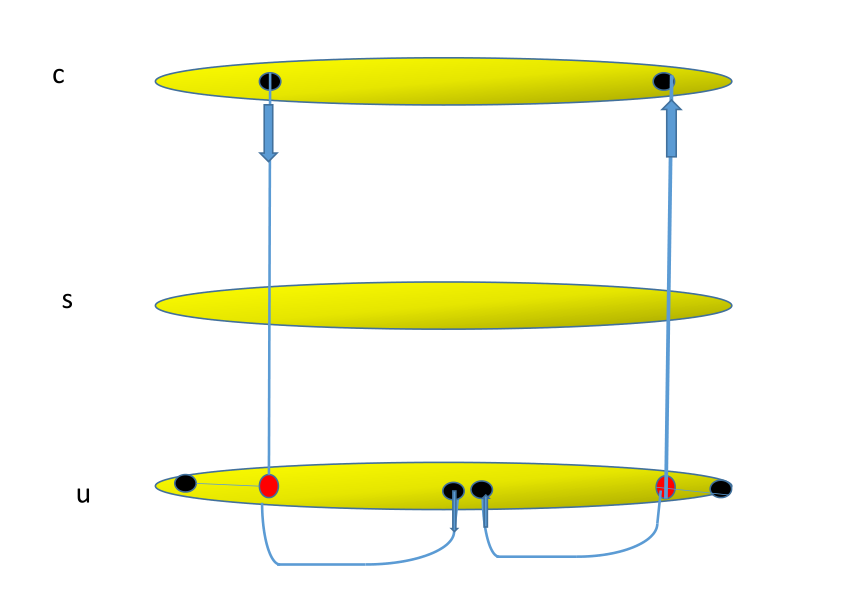}
				\caption{\label{holdecay} The decay  of a holographic tetraquark composed of a diquark of $c$ and $u$ and an anti diquark of $\overline c$ and $\overline u$ into $\lclc$.}
			\end{figure}

\begin{figure} \centering
	\includegraphics[width=0.48\textwidth]{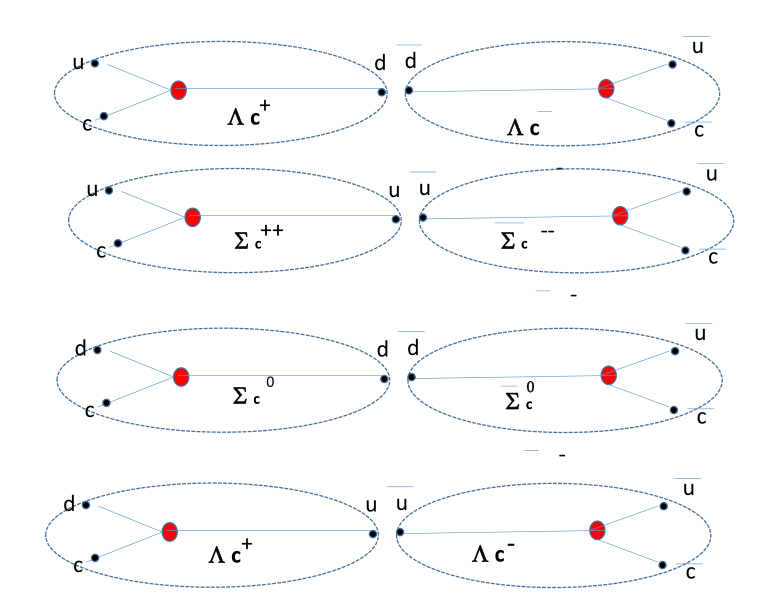}
				\caption{\label{HISHdecay} Different types of tetraquarks based on a diquark of $c$ and $u$ and an anti diquark of $\bar c$ and $\bar u$, and their decay products.}
			\end{figure}

In figure \ref{holdecay} we draw a sketch of the holographic decay of the $Y_{(cd)(\overline{cd})}/Y_{(cu)(\overline{cu})}$ tetraquark into a $\lclc$ pair. Figure \ref{HISHdecay} shows the same process in the HISH description including various other flavor combinations and their outcomes.

It is clear from these figures that the stringy tetraquarks cannot decay through this mechanism of breaking the string into two mesons like $D$ and $\overline D$, but rather only to a baryon-antibaryon pair. On the other hand it is also clear that a stringy charmonium state decays via a string breaking and a creation of quark antiquark pair into a pair of charmed mesons, $D$ and $\overline D$.

The decay width of a tetraquark into a baryon and antibaryon via this mechanism 	has the same structure as that of a meson and a baryon. It was computed in \cite{Peeters:2005fq}, where the decay width is found to be proportional to the length of the string \(l\) times the probability for pair creation,
\be
\Gamma\sim l e^{-\frac{{m^q_{sep}}^2}{T}} \,,
\ee
where $m^q_{sep}$ is the mass of the quark created as an endpoint of the torn apart string. The length is a function of the string tension, the energy, and of $m_1$ and $m_2$, the masses of the diquark and antidiquark at the endpoints of the original tetraquark. For long strings (or small \(m_1\) and \(m_2\)) we have the expression
\be
\Gamma \sim \left(\frac{2 E}{\pi T}-\frac {m_1+m_2}{2T} \right) e^{-\frac{{m^q_{sep}}^2}{T}}\,.
\ee
As was shown in \cite{Peeters:2005fq} the decay probability is proportional to the product of the probability of the horizontal string to reach due to quantum fluctuations a flavor brane (the exponential factor) and the probability that the string splits (which is proportional to the string length).

\section{HISH tetraquarks facing hadron phenomenology} \label{sec:pheno}
In the previous sections we have sketched   several exotic  hadron configurations in the HISH model. Obviously, the most interesting issue is to what extent have such states been observed or could be observed in future experiments. The main idea we would like to introduce here is that tetraquarks should reside on modified Regge trajectories, and that this could an important tool in their phenomenological study.

In this section, we begin by reexamining the trajectories of the regular \(\ccb\) mesons, showing that they fit on trajectories of rotating strings with massive endpoints. We then introduce the \(Y(4630)\), a charmonium-like resonance which is the main tetraquark candidate, and should have a trajectory similar to those of the charmonia. We offer predictions for higher tetraquark states.

Following the discussion of the charmonium sector we discuss the possibility of tetraquarks containing \(\bbb\) or \(\ssb\). In those cases, there are no experimentally  known candidates that decay to a baryon and antibaryon, but we offer some predictions for tetraquark trajectories analogous to that of the charmonium-like \(Y(4630)\).

All the fits in this section were done using the formulae obtained from the rotating string with massive endpoints, given by
\be E = 2m\left(\frac{\beta\arcsin(\beta)+\sqrt{1-\beta^2}}{1-\beta^2}\right)\,, \label{eq:E_sym}\ee

\be J + n - a = 2\pi\alp m^2\frac{\beta^2}{(1-\beta^2)^2}\left(\arcsin(\beta)+\beta\sqrt{1-\beta^2}\right) \,. \label{eq:J_sym}\ee
These are the same equations as those of eqs. \ref{eq:massGenE} and \ref{eq:massGenJ} when we take the symmetric case of two equal endpoint masses. Note that \(J\) is shifted for the quantum case to \(J+n-a\). In this way we also do the radial trajectories of increasing excitation number \(n\).

We note again that, as explained in section \ref{sec:tetra_traj}, the mass (as a string endpoint) of a diquark of a heavy quark and a light one is expected to be the same as the mass of the heavier quark. That is why, to give one example, we take \(m = m_c \approx 1.5\) GeV for both the charmonium and the charmonium-like tetraquark trajectories.

\subsection{Tetraquarks in the charmonium sector}
In recent years, many new charmonium-like resonances have been discovered, and consequently, many papers have been written on the subject. Some useful reviews written in the last few years are \cite{Liu:2013waa,Olsen:2014qna,Esposito:2014rxa,Chen:2016qju}.

\subsubsection{Regge trajectories of the \texorpdfstring{$\ccb$}{c-cbar} mesons} \label{sec:traj_meson}
\begin{figure*} \centering
\includegraphics[width=0.95\textwidth]{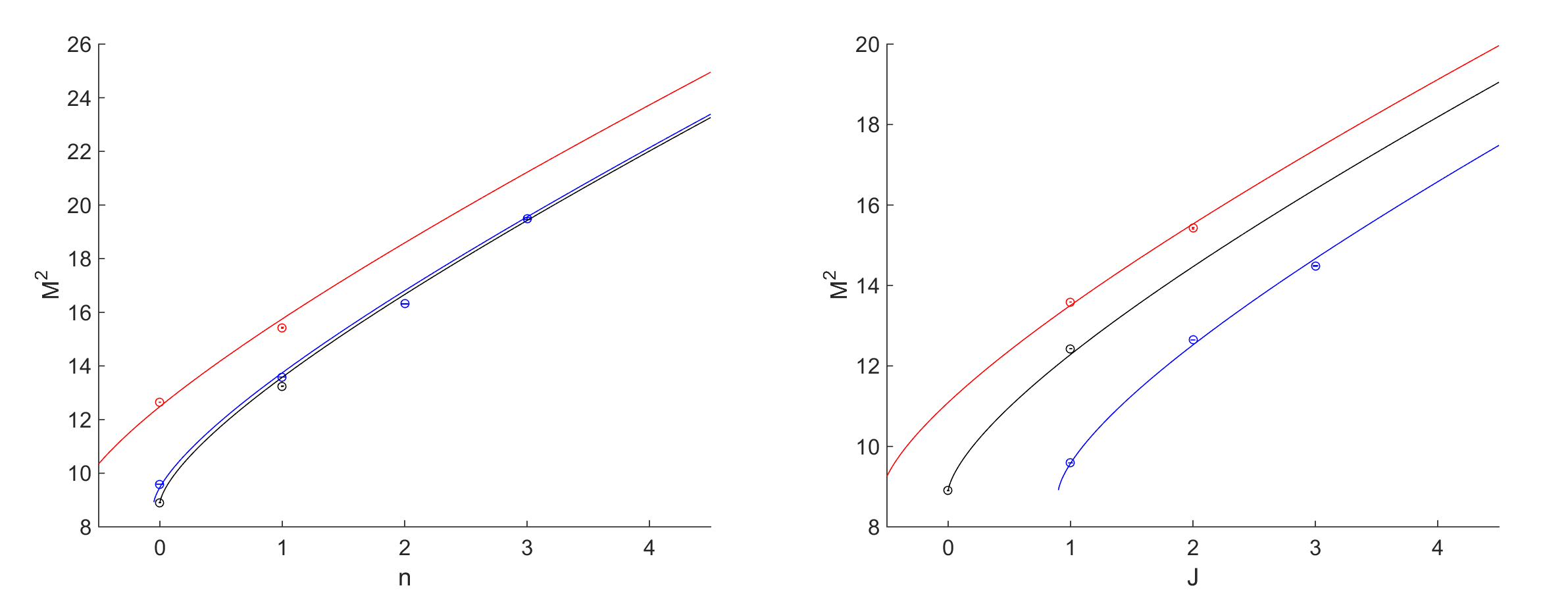}
\caption{\label{fig:traj_meson} Trajectories of \(\ccb\) mesons. \textbf{Left:}  Trajectories in \(n\) of the \(\eta_c\) (black), \(J/\Psi\) (blue), and \(\chi_{c2}\) (red). \textbf{Right:} Orbital trajectories of the \(\eta_c\) (black), \(J/\Psi\) (blue), and \(\Psi(2S)\) (red). See table \ref{tab:traj_meson} for a full specification of the states used.}
\end{figure*}

In \cite{Sonnenschein:2014jwa}, we presented the fits to the trajectories of the \(J/\Psi\) meson, including excited \(\Psi\) mesons and the \(\chi_c\) states. We can expand on the analysis there by adding the trajectories of all of the other \(\ccb\) mesons, namely the trajectories of the different \(\eta_c\) and \(h_c\) states. We take a mass of \(m_{sep} = m_c = 1490\) MeV for the \(c\) quark at the string endpoints. The Regge slope is found to be different for the different types of trajectories. For the orbital trajectories, the slope is 0.86 \GEV\!,\footnote{Note that 0.86 \GEV is almost the same slope one gets for the light mesons, so, at least in the \((J,M^2)\) plane, we have a unified description of the trajectories of all mesons comprised of \(u\), \(d\), \(s\), and \(c\) quarks.} while for the radial trajectories, the slope is 0.59 \GEV. We present all the \(\ccb\) trajectories in figure \ref{fig:traj_meson}. A detailed table of the states used and the masses obtained, including predictions of higher charmonia, is in table \ref{tab:traj_meson}.

In sorting the \(\ccb\) mesons into trajectories, we do not stray from the accepted assignments of the different states. Our predictions are also consistent with those obtained by potential models \cite{Barnes:2005pb}. One omission on the part of our model is that we do not offer a way of predicting the spin-orbit splitting measured for the \(\chi_{cJ}\) and higher spin \(\Psi\) mesons. Therefore in table \ref{tab:traj_meson} we give a single predicted mass for these triplets of states.

\begin{table*}
	\begin{tabular}{|c|c|c|c|c|c|c|c|c|c|c|} \hline
	
		\(J^{PC}\) & \(n\) & State & \(M\) (Exp.) & \(M\) (Thry.) &
		\qquad & \(n\) & \(J^{PC}\) & State & \(M\) (Exp.) & \(M\) (Thry.) \\ \hline\hline
		
		\(0^{-+}\) & 0		&	\(\eta_c(1S)\)   	& 2983.6\plm0.6	& 2980			&&
		0			& \(0^{-+}\)& \(\eta_c(1S)\)		& 2983.6\plm0.6	&	2986			\\ \cline{2-5}\cline{8-11}
		
							 & 1		& \(\eta_c(2S)\)		&	3639.2\plm1.2		&	3647				&&
					& \(1^{+-}\)& \(h_c(1P)\)				&	3525.4\plm0.1		&	3505		\\ \cline{2-5}\cline{8-11}
							
							 & 2		& \(\eta_c(3S)\)		&	?					&	4023				&&
					& \(2^{-+}\) & \(\eta_{c2}(1D)\) & ?					&	3804		\\ \cline{1-5}\cline{7-11}
							
		\(1^{--}\) & 0    & \(J/\Psi(1S)\)	&	3096.9\plm0.0			&	3104						&&
		0			& \(1^{--}\)& \(J/\Psi(1S)\)	& 3096.9\plm0.0			&	3097		\\ \cline{2-5}\cline{8-11}
		
							 & 1		& \(\Psi(2S)\) 		&	3686.1\plm0.0					&	3680						&&
			& \(0^{++}\)& \(\chi_{c0}(1P)\)	  & 3414.8\plm0.3					&							\\ \cline{2-5}

							 & 2		& \(\Psi(3S)\) 		&	4039\plm1			&	4049						&&
			& \(1^{++}\)& \(\chi_{c1}(1P)\)	& 	3510.7\plm0.1	&	3539						\\ \cline{2-5}

							 & 3		&	\(\Psi(4S)\)		&	4421\plm4					&	4354						&&
			& \(2^{++}\)& \(\chi_{c2}(1P)\)	 & 	3556.2\plm0.1 &							\\ \cline{2-5}\cline{8-11}

							 & 4		& \(\Psi(5S)\) 		&		?				&	4621			&&
 & \(1^{--}\)& \(\Psi(1D)\)	 & 3773.1\plm0.3	&					\\ \cline{1-5}
																					
		\(1^{+-}\) & 0		&	\(h_c(1P)\)   		& 3525.4\plm0.1		& [3525]			&&
& \(2^{--}\)& \(\Psi_2(1D)\)	 & ?					&	3830						\\ \cline{2-5}

							 & 1		& \(h_c(2P)\)				&	?					&	3932						&&
& \(3^{--}\)& \(\Psi_3(1D)\)	 & ?					&							\\ \cline{1-5}\cline{8-11}
							
							 \(0^{++}\) & 0		&	\(\chi_{c0}(1P)\) & 3414.8\plm0.3				& [3415]						&& 
& \(j^{++}{}^{[a]}\) & \(\chi_{cj}(1F)\)	 & ?					&	4072				\\ \cline{2-5}\cline{7-11}
	
			 & 1		& \(\chi_{c0}(2P)\)	&	?					&	3854						&&
1 & \(0^{-+}\)& \(\eta_c(2S)\)	 & 3639.2\plm1.2	&	[3639]					\\ \cline{1-5}\cline{8-11}

\(1^{++}\) & 0		&	\(\chi_{c1}(1P)\) & 3510.7\plm0.1	& [3511]						&&
 & \(1^{+-}\)& \(h_c(2P)\)	 & ? 					&	3911			\\ \cline{2-5}\cline{7-11}
							
							& 1		& \(\chi_{c1}(2P)\)	&	?					&	3922						&&
1 & \(1^{--}\) & \(\Psi(2S)\) & 3686.1\plm0.0	& 3676         \\ \cline{1-5}\cline{8-11}
		
\(2^{++}\) & 0		& \(\chi_{c2}(1P)\) &	3556.2\plm0.1					&	3545						&&
 & \(0^{++}\)& \(\chi_{c0}(2P)\)	& ?					&							\\ \cline{2-5}
							
& 1		& \(\chi_{c2}(2P)\) & 3927.2\plm2.6					&	3946		&&
 & \(1^{++}\)& \(\chi_{c1}(2P)\)	 & ?					&	3941			\\ \cline{2-5}
							 
							& 2    & \(\chi_{c2}(3P)\) & ?			&	4266			 &&
 & \(2^{++}\)& \(\chi_{c2}(2P)\)	 & 3927.2\plm2.6 &							\\ \cline{8-11}

			& 		& 	&						&							&&
& \(j^{--}{}^{[b]}\)& \(\Psi_j(2D)\)	 & ?					&	4169				\\ \hline

	\end{tabular}
	\caption{\label{tab:traj_meson} The \(\ccb\) mesons sorted in modified  Regge trajectories, including predictions for higher charmonia. The columns on the left show the radial trajectories of increasing \(n\), while on the right we have trajectories in the (orbital) angular momentum. For each trajectory we compared measured to experimental masses and offer one higher predicted state. The square brackets around some of the predicted masses indicate that the respective states were not used in the fits (and are therefore not drawn in figure \ref{fig:traj_meson}), because we lack the measurements for the next states in their trajectories. Their masses are then used as input to predict their excited partners. [a] \(j=2,3,4\). [b] \(j=1,2,3\).}
\end{table*}

The trajectories of the hidden charm tetraquarks are expected to have the same values of the slopes and endpoint masses as the \(\ccb\) trajectories, so we will continue to use the values listed above in the following section.

\subsubsection{The \texorpdfstring{$Y(4630)$}{Y(4630)} resonance: an overview}
Of particular interest to us is the state \(Y(4630)\). It was observed by Belle in 2008, in the process \(e^+e^-\to \gamma \Lambda^+_c\Lambda^-_c\), with a significance of \(8\sigma\) \cite{Pakhlova:2008vn}. The parameters measured there for the \(Y(4630)\) were
\be J^{PC} = 1^{--}\,,\,\, M_{Y(4630)} = 4634^{+9}_{-11}\,, \,\, \Gamma_{Y(4630)} = 92^{+41}_{-32} \,.\ee

As described in section \ref{sec:decays}, the decay to a baryon-antibaryon pair is the natural decay mode of a tetraquark. We reach this conclusion from a string theory point of view, but this possibility was also explored in other contexts elsewhere \cite{Cotugno:2009ys,Brodsky:2014xia,Liu:2016sip,Guo:2016iej}.

Near the \(Y(4630)\), there is another reported resonance, the \(Y(4660)\). It was observed by both Belle \cite{Wang:2007ea} and BaBar \cite{Aubert:2007zz}, also has \(J^{PC} = 1^{--}\), and its measured mass and width are
\be M_{Y(4660)} = 4665\pm10\,, \qquad \Gamma_{Y(4660)} = 53\pm16\,.\footnote{These are the average values given by the PDG \cite{PDG:2014}.} \ee
The channel in which it was observed is \(e^+e^-\to\gamma\pi^+\pi^-\Psi(2S)\).

There are claims, and possibly even a consensus, that the two resonances \(Y(4630)\) and \(Y(4660)\) are really one and the same, so what we have is a single state with both decay channels, \(\Lambda_c\overline{\Lambda}_c\) and \(\Psi(2S)\pi^+\pi^-\). If this is the case, then the \(\Lambda_c\bar{\Lambda}_c\) is still the dominant decay, so this does not interfere so much with the tetraquark picture \cite{Cotugno:2009ys,Guo:2010tk}.

Other works discussing the \(Y(4630)\) and \(Y(4660)\) include \cite{Ebert:2008kb,Ding:2007rg,Wang:2016fhj}.

\subsubsection{Deciphering the \texorpdfstring{$Y(4630)$}{Y(4630)} state}
Let us now examine the description of the $Y(4630)$ state, the state that decays primarily into \(\Lambda_c\overline{\Lambda}_c\), in terms of possible HISH pictures. There are three possible explanations:

\begin{itemize}

\item A highly excited charmonium state. Specifically, it could be the state \(\Psi(5S)\), as seen in table \ref{tab:traj_meson} (the mass of the \(\Psi(5S)\) should be about 4620 MeV). Such a state should reside on a charmonium Regge trajectory as described in section \ref{sec:traj_meson} and should have heavier higher excited states on the same trajectory, which should also decay to $\Lambda_c\overline{\Lambda}_c$. The decay process involves in this case a creation of a baryonic vertex with a diquark attached to it together with an antibaryonic vertex joined to an antidiquark. Unlike the case of quark-antiquark pair creation where we can estimate in holography the decay width as in \cite{Peeters:2005fq}, the process of creating a pair of baryonic vertices translates in holography to the creation of a wrapped $Dp$ over a $p$-cycle and there is no simple way to estimate its probability.

\item The second option is the tetraquark.In this case too there should be higher excited  tetraquark states residing on a trajectory with the Y(4630) state. The decay process into a pair of $\Lambda_c\overline{\Lambda}_c$ is the one described in section \ref{sec:decays}. The width for such a decay is the same as the decay width of a stringy meson into two stringy meson  s via the breaking of the string and creation of a pair of endpoints.

\item There is also an option which does not take the form of an exotic hadron but rather a molecule of two hadrons. Since a molecule is not a single string with certain dressings on its endpoints and hence we do not expect in such a case to have a trajectory of such states. The simplest possibility is a molecule of $\Lambda_c$ and $\overline{\Lambda}_c$, which then simply breaks apart into the baryon antibaryon pair. The measured mass of the state $Y(4630)$ is higher than twice the $\Lambda_c$ mass by some 60 MeV, when the uncertainty in the measurement of the mass is of about 10 MeV. Had the molecule been bounded with some non-trivial binding energy, its mass should have been smaller than twice the mass of $\Lambda_c$. In \cite{Lee:2011rka} the binding energy was evaluated, in a calculation where it arose from \(\sigma\) or \(\omega\) exchange. Here there is no such binding energy involved. The \(\Lambda_c\bar\Lambda_c\) molecule is then disfavored, but not entirely ruled out given its large measured width of  \(92^{+41}_{-32}\) MeV. On the other hand, the \(Y(4630)\) and the \(Y(4660)\) resonance mentioned in the previous section are near, and slightly below, another threshold, \(\Psi(2S)f_0(980)\), which is at \(4676\pm20\) MeV. According to \cite{Guo:2010tk}, the interpretation of both measurements as a single resonance with this molecular configuration is not inconsistent with experimental results, and in particular can reproduce the \(\Lambda_c\bar\Lambda_c\) decay width. In the stringy picture, the decay of the proposed \(\Psi(2S)f_0(980)\) molecule into \(\Lambda_c\bar\Lambda_c\) would again require the creation of a BV-anti-BV pair, and its probability difficult to estimate.

\end{itemize} 

To summarize, the current data regarding the \(Y(4630)\) resonance is not enough to determine its nature. This is why we propose to search for more states on its Regge trajectory. Its excited partners, which, if found, will serve as evidence for the tetraquark picture, and in particular could make it preferable to the molecular configuration, which does not exhibit trajectories.

\subsubsection{Regge trajectory of the \texorpdfstring{$Y(4630)$}{Y(4630)}} \label{sec:y4360}
If the \(Y(4630)\) is indeed a tetraquark, it should be part of a Regge trajectory. Using the values of \(m_{sep}\) and \(\alp\) of the \(\ccb\) trajectories, we can use the known mass of the state \(Y(4630)\) and extrapolate from it to higher (or lower) states along the trajectory. From the combined fit of section \ref{sec:traj_meson}, we have the values
\be m_c = 1490 \MEV\,, \ee
and
\be \alp_J = 0.86 \GEVm\,,\qquad \alp_n = 0.59 \GEVm\,. \ee
There are two different slopes, one for orbital trajectories and one for radial trajectories.

The higher states are only predictions at this point, as measurements at the relevant masses are scarce, but it is interesting to see if we can continue the trajectory backwards, that is, to check if the \(Y(4630)\) is an excited state of an already known resonance. The lower state would be below the \(\Lambda_c\bar{\Lambda}_c\) threshold, and its decay would require the baryonic and antibaryonic vertices to annihilate. This could mean that such a state is very narrow, due to a OZI-like suppression mechanism of these decays. This mechanism was discussed recently in \cite{Rossi:2016szw}.

In the angular momentum plane, we can have a scalar state with \(J^{PC} = 0^{++}\) preceding the \(Y(4630)\) in its trajectory. There is no currently known scalar resonance at the mass predicted by the Regge trajectory, of about 4480 MeV.

In the \((n,M^2)\) plane we may find the \(Y(4360)\), another vector resonance that is at the right mass for the \(Y(4630)\) to be its first excited state. Its mass and width are
\be M_{Y(4360)} = 4354\pm10\,, \qquad \Gamma_{Y(4360)} = 78\pm16\,. \ee
It was observed in decays to \(\Psi(2S)\pi^+\pi^-\). This is the same decay channel in which the \(Y(4660)\) was observed. In fact, suggestions that \(Y(4630)\) or \(Y(4660)\) is an excited partner of \(Y(4360)\) are not uncommon \cite{Guo:2010tk}.

The problem with assigning the \(Y(4360)\) as the unexcited partner of the \(Y(4630)\) is in its width. The widths of the two states are roughly the same, even though the lower state, being beneath the \(\Lambda_c\bar{\Lambda}_c\) threshold, should be significantly narrower than the above threshold state.

Here we also have to reconsider the \(Y(4660)\), as it was observed in the same channel as the \(Y(4360)\). One scenario exists in which the \(Y(4630)\) and \(Y(4660)\) are the same resonance, that decays both to \(\Lambda_c\bar{\Lambda}_c\) and \(\Psi(2S)\pi^+\pi^-\). It is a tetraquark, but there is a limited phase space for \(\Lambda_c\bar{\Lambda}_c\) decays, and we see it decay to \(\Psi(2S)\pi^+\pi^-\). The \(Y(4360)\) is also a tetraquark, but has no phase space to decay into baryons and decays instead to \(\Psi(2S)\pi^+\pi^-\). However, we have to ask why the tetraquark would have such a large decay width for non-baryonic channels. We will leave this scenario for now, and consider the higher excitations of \(Y(4630)\).

Whether or not there is a lower, unexcited version of \(Y(4630)\), we still predict higher excitations of the stringy tetraquark. For the higher states, the \(\Lambda_c\bar{\Lambda}_c\) decays should be dominant, and their masses should be such that they fall on a Regge trajectory. The predictions are in table \ref{tab:pred_y_c}.

The intercept of the Regge trajectories is calculated by matching the mass formula of eqs. \ref{eq:E_sym} and \ref{eq:J_sym} to the mass of the \(Y(4630)\). The values obtained are \(a = -5.8\) for the orbital trajectory, and \(a = -4.0\) for the radial trajectory. If we want to test this against the formula conjectured in section \ref{sec:quantum_traj}, where the intercept of the tetraquark is 
\be \tilde a_t = \tilde a + 2 \Delta \tilde a \,.\ee
The value \(\tilde a\) is the intercept of the meson with the same flavor content, in this case the charmonium. We take the intercept of the \(J/\Psi\) trajectory, which is very small for both the radial and orbital trajectories and equals \(\approx-0.1\). Now \(\Delta \tilde a\) is the difference in the intercept when replacing a quark with a diquark at one of the string's endpoints.\footnote{To be exact, the replacement is of an antiquark with a diquark, or a quark with an antidiquark.} We would like to evaluate it in a case where a \(c\) quark is involved. We could compare the \(J/\Psi\) intercept to the intercept obtained from the mass of the doubly-charmed baryon \(\Xi_c\). From these we have \(\Delta \tilde a\) in the range -0.7 to -1.0, depending of the value of the slope. Another measurement would compare the intercept of charmed mesons (\(D\)) and baryons (\(\Lambda_c\)). Then we get \(\Delta \tilde a = (-0.7)\)--\((-0.5)\). These values obviously do not satisfy the equation when \(\tilde a_t\) is as negative as we found it to be.

This reflects the large mass difference between the \(Y(4630)\) state and the \(\ccb\) ground state mass, and is perhaps an indication that the \(Y(4630)\) is a radially excited state and not the first state in its trajectory. If we assume that it has \(n = 2\), for example, instead of \(n = 0\), then the intercept of its radial trajectory is \(-2.0\), and we get a better agreement with the theory provided there are two more states in the trajectory beneath the \(Y(4630)\).

\begin{table}[h!] \centering
	\begin{tabular}{|c|c|c|} \hline
\(n\)	&	 Mass 	&	 Width \\ \hline\hline
``-2'' & 4060\plm50 & Narrow \\ \hline
``-1'' & 4360\plm50 & Narrow \\ \hline\hline
0 & \(\bf{4634^{+9}_{-11}}\)	&	\(\bf{92^{+41}_{-32}}\)	\\ \hline\hline
1 &	4870\plm50 	&	 150--250	\\ \hline
2	&	5100\plm60 	&	 200--300	\\ \hline
3	&	5305\plm60 	&	 220--320	\\ \hline
4	&	5500\plm60 	&	 250--350	\\ \hline
\end{tabular} \qquad\qquad
\begin{tabular}{|c|c|c|} 
\multicolumn{3}{c}{} \\ \hline
\(J^{PC}\) &	  Mass 	&	 Width \\ \hline\hline
\(0^{++}\) & 4485\plm40					& Narrow \\ \hline\hline
\(1^{--}\) & \(\bf{4634^{+9}_{-11}}\)					& \(\bf{92^{+41}_{-32}}\)	\\ \hline\hline
\(2^{++}\) & 4800\plm40 	&	 150--250	\\ \hline
\(3^{--}\) & 4960\plm40 	&	 180--280	\\ \hline
\(4^{++}\) & 5100\plm45 	&	 200--300	\\ \hline
\(5^{--}\) & 5260\plm45 	&	 250--350	\\ \hline
\end{tabular} \caption{\label{tab:pred_y_c} Trajectories of the \(Y(4630)\). Based on the experimental mass and width of the \(Y(4630)\) we extrapolate to higher excited states on the trajectory. Uncertainties are based on both experimental errors and uncertainties in the fit parameters. The excited states are expected to decay into \(\Lambda_c\overline{\Lambda}_c\). Some possible lower states, with masses below the \(\Lambda_c\overline{\Lambda}_c\) threshold, are also included. \(Y(4360)\), observed to decay to \(\Psi(2S)\pi^+\pi^-\), is a candidate for the \(n = -1\) state. See text for details.}
\end{table}

\subsection{Tetraquarks in the bottomonium sector}
In the spectroscopy of heavy quarkonia, one usually expects analogies to exist between the \(\ccb\) and \(\bbb\) spectra. Tetraquarks and other exotics should be no exception. Therefore, we propose to search for an analogous state to the \(Y(4630)\), which we will call the \(Y_b\), that decays primarily to \(\Lambda_b\bar{\Lambda}_b\). We will assume that, like the \(Y(4630)\), the \(Y_b\) state is located a little above the relevant baryon-antibaryon threshold of \(\Lambda_b\overline{\Lambda}_b\).

The mass of the \(\Lambda_b^0\) is \(5619.51\pm0.23\) MeV. As an estimate - take a mass of
\be \begin{split} & M(Y_b) \approx 2M(\Lambda_b^0)+40\MEV \\
 & \Delta M(Y_b) \approx 40\MEV \end{split} \ee
The \(Y_b\) would also have its own Regge trajectory, so, based on the mass we choose for it, we can predict the rest of the states that lie on the trajectory. The trajectories are calculated using the slopes of the \(\Upsilon\) trajectories, which are \(\alp \approx 0.64\GEVm\) in the \((J,M^2)\) plane, and \(\alp\approx0.46\GEVm\) in the \((n,M^2)\) plane. The string endpoint mass of the \(b\) quark is \(m_b = 4730\) MeV.\footnote{Values taken from the Regge trajectory fits of \cite{Sonnenschein:2014jwa}.} The resulting masses are listed in table \ref{tab:pred_y_b}.

Like before, we get a large negative intercept if we assume the above threshold state \(Y_b\) is the first state in the trajectory with \(n = 0\). For the radial trajectory, this intercept is -6.1. So, we may also continue the radial trajectory backwards. The masses of the states that would preceed the above threshold \(Y_b\) tetraquark are also listed in table \ref{tab:pred_y_b}. We can see that the observed resonance \(Y_b(10890)\) is a potential match to be the ``\(n = -2\)'' state. This resonance was observed to decay to \(\pi^+\pi^-\Upsilon(nS)\) and has a mass of \(10888.4\pm0.3\) \cite{Olsen:2014qna,Abe:2007tk,Ali:2009pi}. It seems analogous to the \(Y(4360)\) from the charmonium sector, which decayed to \(\pi^+\pi^-\Psi(2S)\) and was at the right mass to be an unexcited partner of the tetraquark candidate \(Y(4630)\). If the \(Y(10890)\) is on the trajectory, then we also predict another resonance following it, which being similarly below the baryon-antibaryon threshold should have similar properties, at a mass of \(\approx11080\) MeV.

\begin{table}[h!] \centering
	\begin{tabular}{|c|c|} \hline
\(n\)	&	 Mass 	\\ \hline\hline
``-2'' & 10870\plm50 \\ \hline
``-1'' & 11080\plm50 \\ \hline\hline
0 & 11280\plm40	\\ \hline
1 &	11460\plm40 \\ \hline
2	&	11640\plm40 \\ \hline
3	&	11810\plm40 \\ \hline
4	&	11980\plm40 \\ \hline
\end{tabular} \qquad\qquad
\begin{tabular}{|c|c|} \hline
\(J^{PC}\) &	  Mass 	\\ \hline\hline
\(1^{--}\) & 11280\plm40	\\ \hline
\(2^{++}\) & 11410\plm40 	\\ \hline
\(3^{--}\) & 11550\plm40 	\\ \hline
\(4^{++}\) & 11670\plm40	\\ \hline
\(5^{--}\) & 11800\plm40 	\\ \hline
\end{tabular} 
\caption{\label{tab:pred_y_b} Predictions for the states of the \(Y_b\), a tetraquark containing \(\bbb\) and decaying to \(\Lambda_b\bar{\Lambda}_b\). The mass of the first state is taken near threshold, masses of higher states are on the Regge trajectories that follow from the ground state.}
\end{table}

\subsection{Lighter tetraquarks}
An important and longstanding open question is why there are no tetraquarks, or strong tetraquark candidates among the light mesons. In this note we will not attempt to answer this question, even though among the 225 possible tetraquarks we counted section \ref{sec:tetra}, many are made up of \(u\), \(d\), and \(s\) quarks and their antiquarks only.

Instead, suppose that at least \(s\) quarks are heavy enough to accommodate tetraquarks. Then we again predict a state that would decay to \(\Lambda\bar{\Lambda}\), and a trajectory of its excited states. We then predict again a trajectory of states beginning with a near-threshold state. The masses are in table \ref{tab:pred_y_s}.

For the predictions, we take a ground state mass of
\be M(Y_s) \approx (2M(\Lambda)+40)\pm40\MEV \ee
with the usual slope for light quark (\(u\), \(d\), \(s\)) hadrons of 0.9 \GEV in the \((J,M^2)\) plane. In the \((n,M^2)\) the slope is lower, and we take 0.8 \GEV. The mass of the \(s\) quark is taken to be 220 MeV.

The resulting intercept is -3.6. Again we repeat the exercise of extrapolating the trajectory backwards, listing two preceding states. The vector resonance \(\rho(1570)\) might be a match.

\begin{table}[h!] \centering
	\begin{tabular}{|c|c|} \hline
\(n\)	&	 Mass 	\\ \hline\hline
``-2'' & 1580\plm40	\\ \hline
``-1'' & 1960\plm40	\\ \hline\hline
0 & 2270\plm40	\\ \hline
1 &	2540\plm40 		\\ \hline
2	&	2780\plm40 		\\ \hline
3	&	3000\plm40 		\\ \hline
4	&	3210\plm40 		\\ \hline
\end{tabular} \qquad\qquad
\begin{tabular}{|c|c|} \hline
\(J^{PC}\) &	  Mass 	 \\ \hline\hline
\(1^{--}\) & 2270\plm40			\\ \hline
\(2^{++}\) & 2510\plm40 		\\ \hline
\(3^{--}\) & 2730\plm40 		\\ \hline
\(4^{++}\) & 2930\plm40			\\ \hline
\(5^{--}\) & 3120\plm40 		\\ \hline
\end{tabular} 

\caption{\label{tab:pred_y_s} Predictions for the states of the \(Y_s\), a tetraquark containing \(\ssb\) and decaying to \(\Lambda\bar{\Lambda}\).}
\end{table}
\section{Summary} \label{sec:summary}
In this note we have used the HISH model to shed some light on the tetraquark controversy, and we used recent data on the spectra and decays of  heavy hadrons as a further test of the HISH model.
The main theme of this note is quite simple. Taking as a definition that a hadron is a string (or a connected collection of strings), then any exotic hadron, just as the normal hadrons, has to admit a Regge trajectory behavior in its spectrum. Conversely, this is not the case for a ``molecule'', or bound state of strings. Thus, a simple way to verify that a state is indeed a multiquark exotic hadron is to look for its excited Regge partners. Of course, the absence of these excited partners on a trajectory cannot be considered conclusive evidence that a state is not a tetraquark. However, the existence of a trajectory would strongly imply that the state is a tetraquark.

We suggest to apply this rule to the resonance Y(4630). Due to the fact that it predominantly decays to a baryon and an antibaryon, it seems plausible from its HISH model description that it is indeed a genuine stringy tetraquark, and, if we consider the Regge trajectory, it is a possible gateway leading to further excited tetraquark states. We further propose to look for a similar phenomena also for hadrons containing a $b$ quark and a $\bar b$, as well as $s$ and $\bar s$, also decaying to baryon-antibaryon. We recap the predictions for these tetraquarks in table \ref{tab:pred_summary}.

\begin{table*} \centering
	\begin{tabular}{|c|c|c|c||c|c|c||c|c|c|} \hline
	\multicolumn{4}{|c||}{Tetraquarks containing \(\ccb\)}	& \multicolumn{3}{|c||}{Tetraquarks containing \(\bbb\)} & \multicolumn{3}{|c|}{Tetraquarks containing \(\ssb\)} \\
	\multicolumn{4}{|c||}{(decaying to \(\lclc\))}	& \multicolumn{3}{|c||}{(decaying to \(\Lambda_b\overline\Lambda_b\))} & \multicolumn{3}{|c|}{(decaying to \(\Lambda\overline\Lambda\))} \\ \hline
	
	\(n\) & \(J^{PC}\) & \(M\) & \(\Gamma\) & \(n\) & \(J^{PC}\) & \(M\) & \(n\) & \(J^{PC}\) & \(M\) \\ \hline
	
	0 & \(1^{--}\) & \(\bf{4634^{+9}_{-11}}\) & \(\bf{92^{+41}_{-32}}\) &
	0 & \(1^{--}\) & 11280\plm40 &
	0 & \(1^{--}\) & 2270\plm40 \\ \hline
	
	0 & \(2^{++}\) & 4800\plm40 & 150--250 & 
	0 & \(2^{++}\) & 11410\plm40 & 
	0 & \(2^{++}\) & 2510\plm40 \\ \hline
	
	1 & \(1^{--}\) & 4870\plm50 & 150--250 & 
	1 & \(1^{--}\) & 11460\plm40 & 
	1 & \(1^{--}\) & 2540\plm40 \\ \hline
	
	0 & \(3^{--}\) & 4960\plm40 & 180--280 & 
	0 & \(3^{--}\) & 11550\plm40 & 
	0 & \(3^{--}\) & 2730\plm40 \\ \hline
	
	2 & \(1^{--}\) & 5100\plm60 & 200--300 & 
	2 & \(1^{--}\) & 11640\plm40 & 
	2 & \(1^{--}\) & 2780\plm40 \\ \hline	
	\end{tabular} \caption{\label{tab:pred_summary} Summary of predictions for tetraquarks and their first few excited states. Higher excited states, and possible lower unexcited ones (which are below the relevant baryon-antibaryon threshold), were listed separately in tables \ref{tab:pred_y_c}, \ref{tab:pred_y_b}, \ref{tab:pred_y_s}. In addition, we made predictions for higher charmonia, which were listed in table \ref{tab:traj_meson}. For the charmonium like tetraquark we estimate the width based on the width of the measured state \(Y(4630)\) (here in bold).}
\end{table*}

We note that while the natural way to search for molecules is by looking in the vicinity of different two hadron thresholds \cite{Karliner:2015ina}, a good way to identify tetraquarks is to look above the two baryon threshold. It is there that we can look for particles that decay into a baryon-antibaryon pair, the stringy tetraquark's natural and distinctive mode of decay, but it is also a region where data is still quite scarce. And while the predictions above are for the \(Y(4630)\), and its immediate analogues \(Y_b\) and \(Y_s\), it should be noted that there could be similar states with non-symmetric decay channels, \(\Lambda_c\overline\Lambda_b\) for example. For resonances found in such channels we will also expect trajectories, which can be easily predicted using the HISH model.

There are several future directions along which this research project can be continued. Here we list some of them.
\begin{itemize}
\item
While the focus here was on the $Y(4630)$ state, as a window to the exotic hadrons, there are in fact there are plenty of other exotic resonances in the charmonium and bottomonium sectors, the plethora of $X$, $Y$ and $Z$ resonances, that are awaiting to be part of the HISH community. They are mostly lower mass states, below the baryon-antibaryon threshold, but some of these resonances may turn out to be themselves tetraquarks. Our model easily describes excited states, with a single long string stretched between diquark and antidiquark, but one should also think about tetraquarks where all four constituent quarks and antiquarks are close together. The latter picture is the one that might be more appropriate to already discovered states, which do not decay to baryon-antibaryon (and without being narrow as a result). This is related to the difficult theoretical question of the applicability of these stringy models to the hadron spectrum. As we have pointed out, the lowest spin mesons and baryons belong on Regge trajectories, showing a stringy nature even when effective long string approximations are expected to break down. One must ask then to what extent this is also true of tetraquarks and other exotics.
\item
As was described in section \ref{sec:exoticHISH} hypothetically there could be 225 types of tetraquarks, which include symmetric, semisymmetric and asymmetric ones. In this note we discussed those that are chargeless and hence decay to a baryon and its antiparticle but there could be also those that carry charges of 1 or even 2. We intend in the future to provide theoretical predictions about their masses and in certain cases also their width. In addition, there are more options to get tetraquarks on top of that built from one BV and one anti-BV. The configuration drawn in figure \ref{multiquark} with two BVs and two anti-BVs is one such example.
\item
Moreover, it is easy to visualize higher order multiquark states like pentaquarks (see figure \ref{molecule}) heptaquarks, and so on. In a similar manner to the description of the MRT of the tetraquarks we will analyze in the future also the trajectories of potential higher multiquarks. 
\item
In this note we addressed the issues of the spectra and decay modes of candidates of exotic quarks. We have not discussed the question of production mechanisms of such states. This issue is obviously crucial for the future generation of exotic hadrons. One should separately discuss the production processes in $e^+e^-$ colliders and in hadron colliders.
\item
From the theoretical side of the story, it is essential to understand better the quantum corrections of the trajectories, namely the intercept, and its generalization, in particular for hadrons that include baryonic vertices. As was mentioned in section \ref{sec:quantum_traj}, that is in fact also the understanding of the difference between meson and baryon trajectories.
\item
So far we used the HISH model to describe genuine exotic hadrons but one should be able to address in this framework also states which are molecules of hadrons both mesons and baryons. Accomplishing this task will shed more light of how to distinguish between the molecule and the exotic hadrons and in addition may be useful for the understanding of molecules, the most famous of which is probably the deuterium.
\end{itemize}

\section*{Acknowledgments}
 We are grateful to O. Aharony, V. Kaplunovsky,  M. Karliner, S. Nussinov,  A. Soffer and S. Yankielowicz for insightful conversations. We want to thank O. Aharony for his remarks  on the manuscript. J.S would to thank Hai-Bo Li of the BESIII collaboration for hosting him  in the IHEP Beijing where part of this work was done. This work was supported in part by a center of excellence supported by the Israel Science Foundation (grant number 1989/14), and by the US-Israel bi-national fund (BSF) grant number 2012383 and the Germany Israel bi-national fund GIF grant number I-244-303.7-2013.

\bibliographystyle{elsarticle-num}
\bibliography{tetra}

\end{document}